\documentclass[12pt]{article}

\usepackage{soul} 

\usepackage{epsfig}

\begin{document}

\begin{center}

{\Large CLIFFORD TORI AND UNBIASED VECTORS}

\vspace{18mm}

{\large Ole Andersson}\footnote{ole.andersson@fysik.su.se}

\

\

{\large Ingemar Bengtsson}\footnote{ibeng@fysik.su.se}

\vspace{15mm}

{\sl Fysikum, Stockholms Universitet,}

{\sl S-106 91 Stockholm, Sweden}

\vspace{12mm}

{\bf Abstract:}

\end{center}

\

\noindent The existence problem for mutually unbiased bases is an unsolved problem 
in quantum information theory. A related question is whether every pair of bases 
admits vectors that are unbiased to both. Mathematically this translates to the 
question whether two Lagrangian Clifford tori intersect, and a body of results 
exists concerning it. These results are however rather weak 
from the point of view of the first problem. We make a detailed study of how the 
intersections behave in the simplest non-trivial case, that of complex projective 
2-space (the qutrit), for which the set of pairs of Clifford tori can be usefully 
parametrized by the unistochastic subset of Birkhoff's polytope. Pairs that do not 
intersect transversally are located. 
Some calculations in higher dimensions are included to see which results are 
special to the qutrit.

\

\vspace{12mm}


Birkhoff's polytope

\newpage

{\bf 1. Introduction}

\hspace{5mm}

\noindent The existence problem for Mutually Unbiased Bases (MUB) is easy to state, arises 
naturally when asking questions about the foundations of quantum mechanics, and may 
have various practical implications \cite{DEBZ}. After several decades of work by quantum 
physicists, it remains open. From a mathematical point of view it is 
not clear how to address this problem. 
It just might be bound up with unsolved questions in discrete mathematics 
such as the existence problem for finite projective planes \cite{Lam}, as has been 
discussed in the literature \cite{Paterek,Weigert,Weiner}. As noticed recently it may also 
be related to questions in 
Lagrangian intersection theory, an at first sight unrelated area of mathematics with roots 
in very different soil \cite{Arnold}. 

On the physics side the set of all vectors unbiased with respect to a given basis 
can be regarded as `maximally quantum' states, as seen by an observer having access 
to only one von Neumann measurement. In mathematics this set of vectors is known as a 
Clifford torus, and forms a Lagrangian submanifold of complex projective space. Thus, 
listing the set of all vectors unbiased to two different bases is equivalent to the 
problem of enumerating the intersections between two Clifford tori. Provided the 
intersections are transversal (they may not be) a powerful mathematical theorem 
guarantees that at least $2^{N-1}$ distinct intersections occur when the Hilbert space has 
$N$ dimensions \cite{Biran, Cho}. However, to be of use in the MUB existence problem these intersections 
have to occur in a very special constellation. 

The case $N = 2$ is trivial. In this case a Clifford torus is a great circle on the Bloch sphere. 
Two great circles always intersect in two antipodal points, and these points represent 
an orthonormal basis unbiased to the two bases one started out with. The purpose of this 
paper is to investigate the simplest non-trivial case, $N = 3$, in full detail. Using a natural 
parametrization of the set of all pairs of Clifford tori we find that the number of 
intersections is 4 or 6, with 3 or 5 intersections occurring in exceptional 
non-transversal cases. 

Section 2 briefly reviews the MUB existence problem and the subject of biunimodular 
vectors. Section 3 explains why questions 
about intersecting Lagrangian tori are relevant to it, and recalls some facts about 
intersecting submanifolds. Section 4 parametrizes  
the set of Clifford tori by means of the unistochastic subset of Birkhoff's polytope. 
A detailed description of the $N = 3$ case is given since it has independent interest. 
Sections 5 and 6 present our main result, a description of how the regions with four and 
six intersection points sit inside the parameter space. 
We pay special attention to sets of measure zero 
where the transversality condition fails. 
Most of the story is based on Mathematica calculations, 
with some analytical support. We believe that 
the resulting picture is basically complete---when $N = 3$. The question whether it has 
any implications for the MUB existence problem is deferred to the concluding section 7. 
Some analytical calculations for higher dimensions are given in two appendices, 
to see to what extent our results are special to the $N = 3$ case.

\vspace{10mm}

{\bf 2. The MUB existence problem and biunimodular vectors}

\vspace{5mm}

\noindent In quantum mechanics an orthonormal basis is needed to specify a von Neumann 
measurement. Given the Hilbert space ${\bf C}^N$, two orthonormal bases 
$\{|e_i\rangle \}_{i=1}^N$ and $\{|f_i\rangle \}_{i=1}^N$ are said to be {\it unbiased} if 
all the numbers $|\langle e_i|f_j\rangle |^2$ are the same. 
Thus the transition probabilities are 

\begin{equation} p(i|j) = |\langle e_i|f_j\rangle |^2 = \frac{1}{N} \label{MUB} \end{equation}

\noindent for all $i,j$. Pairs of bases related in this way have practical applications 
to quantum cryptography, and the concept is closely related to Bohr's 
notion of complementarity \cite{Schwinger}, supposedly central to the philosophy 
of quantum mechanics. 
One obtains an example of an unbiased pair for any dimension 
$N$ by letting the first basis be the computational basis, and then acting on its vectors 
by means of the Fourier matrix $F$, with matrix elements 

\begin{equation} F_{jk} = \frac{1}{\sqrt{N}}e^{\frac{2\pi i}{N}jk} 
\ , \hspace{5mm} 0 \leq j,k \leq N-1 \ . \end{equation}

\noindent The resulting basis is called the Fourier basis. 
Its vectors are made up of the columns of the Fourier matrix. 

In the {\it MUB existence problem} one goes on to ask how many mutually unbiased bases 
there can be, given $N$. 
An answer would have implications ranging from practical quantum state 
tomography \cite{Ivanovic, Wootters} and cryptography \cite{Mo, Mafu} 
to pure mathematics \cite{Kostrikin}. It is easy 
to show that the number is bounded from 
above by $N+1$. Group theoretical constructions of considerable elegance do yield $N+1$ 
mutually unbiased bases in ${\bf C}^N$ if the dimension $N$ is a power of a prime 
number (and yield at least 3 bases in every dimension) \cite{Fields, Godsil, Kantor}. The 
reason behind this success has been pinpointed \cite{Vatan, ACW, Zhu}. 
Extensive computer calculations have produced convincing evidence, but no proof, that the 
maximum number for $N = 6$ is 3 \cite{BS, Jam, Raynal}. Not much more is known about 
this problem. 

A less ambitious problem is that of enumerating the set of all vectors that are unbiased 
with respect to both members of a pair of unbiased bases. For the particular case that this 
pair is the computational basis and the Fourier basis, such vectors are also known 
as biunimodular vectors. They have received 
considerable attention 
(also for reasons unrelated to quantum mechanics) \cite{Gosta, Haagerup}. 
See Table \ref{tab:states}, and observe that for $N > 5$ 
the number of such vectors is larger than one would need in order to construct an 
additional set of $N-1$ bases. For $N = 6$ one finds 48 such vectors, and no less than 
16 bases can be constructed from them \cite{Grassl}. But there are no unbiased pairs of 
bases there. This does not settle the MUB existence problem for $N = 6$ as continuous 
families of unbiased pairs not equivalent to this particular one do exist \cite{Bengt, 
Szollosi}. 

\begin{table}[h]
\caption{{\small Number (\#) of vectors unbiased to both the computational and 
Fourier bases in dimension $N$ \cite{Gosta,Haagerup,Grassl}. If the 
dimension is divisible by a square the intersection of the two Clifford tori is not transversal.}}
  \smallskip
  \smallskip
\hskip 1.6cm
{\renewcommand{\arraystretch}{1.1}
\begin{tabular}
{|c| c|c|c|}\hline \hline
$N$ & $2^{N-1}$ & $N(N-1)$  & \# \\
 \hline
2 & 2
         & 2 & 2 \\
3 & 4
         & 6 & 6 \\
4 & 8
          & 12 & $\infty$ \\
5 & 16 & 20 & 20 \\
6 & 32 & 30 & 48 \\
7 & 64 & 42 & 532 \\
\hline \hline
\end{tabular}
}
\label{tab:states}
\end{table} 

\

Finally we can ask for all vectors unbiased with respect to an arbitrary, not necessarily 
unbiased, pair of bases. This is the subject of the next section, albeit in a different 
language. Again the answer has a practical application to do with error-disturbance 
relations in quantum theory \cite{Rudolph, Karol}, an observation that inspired 
this paper. It also has ramifications for the structure of unitary matrices 
\cite{Idel, FR}. The paper by F\"uhr and Rzestotnik begins with an up-to-date and 
very readable review of the subject of biunimodular vectors \cite{FR}.

\vspace{5mm}

{\bf 3. Clifford tori and their intersections}

\vspace{5mm}

\noindent Quantum states are described, not by vectors in ${\bf C}^N$, but by points 
in complex projective space ${\bf CP}^n$ where $n = N-1$. This is a K\"ahler space with 
a natural metric and a compatible symplectic form. 
For $N = 3$ an arbitrary point is described, in 
the computational basis, by the normalized vector 

\begin{equation} \psi = \left( \begin{array}{l} \sqrt{p_0} \\ \sqrt{p_1}e^{i\nu_1} \\ 
\sqrt{p_2}e^{i\nu_2} \end{array} \right) 
, \ \hspace{5mm} 0 \leq \nu_1, \ \nu_2 \leq 2\pi \ , \hspace{5mm}  
p_0+p_1+p_2 = 1 \ . \label{Cliff1} \end{equation}

\noindent The vector $\vec{p}$ is a probability vector. If we use the components $p_1$ and $p_2$ 
as two of the coordinates the symplectic form takes the canonical form 

\begin{equation} \Omega = dp_1 \wedge d\nu_1 + dp_2\wedge d\nu_2 \ . \end{equation}

\noindent We are now using action-angle coordinates, with the phases $\nu_i$ 
parametrizing tori in phase space \cite{Arnold}. Unitary time evolution is simply Hamiltonian time 
evolution, for Hamiltonian functions of the special form \cite{Kibble, Brody}

\begin{equation} H(p, \nu) = \langle \psi |\hat{H}|\psi \rangle \ . \end{equation}

\noindent If $\hat{H}$ is a diagonal matrix the Hamiltonian function $H$ is independent of 
the angle variables. In conclusion, complex projective space is a phase space, and 
unitary transformations are (quite special) canonical transformations. 

A torus is {\it Lagrangian} if it has half the dimension of phase space and if the symplectic 
form vanishes when restricted to the torus. The action variables $p_i$ parametrize a set 
of Lagrangian tori that fill the phase space. The metric induced on such a torus is \cite{Ingemar} 

\begin{equation} ds^2 = p_1(1-p_1)d\nu_1^2 + p_2(1-p_2)d\nu_2^2 - 2p_1p_2d\nu_1d\nu_2 \ . 
\label{torus} \end{equation}

\noindent 
The (metric) area is maximized when the probability vector is flat, that is 
when 

\begin{equation} p_0 = p_1 = p_2 = \frac{1}{3} \ . \end{equation}

\noindent By definition a {\it Clifford torus} is a Lagrangian torus of maximal area. 
From the example it is clear that 
there will be one Clifford torus for each choice of basis in Hilbert space. It is 
also 
clear from eq. (\ref{MUB}) that a vector is unbiased with respect to a given 
basis if and only if it corresponds to a point on the Clifford torus defined by that 
basis. 

Mathematicians have devoted considerable attention to the question when two Lagrangian 
submanifolds, related by a Hamiltonian deformation (in some symplectic manifold), 
intersect. For the particular case of 
Clifford tori in complex projective space the result alluded to in the introduction---that 
the number of distinct intersections is at least $2^n$ whenever the intersections are 
transversal---is available \cite{Biran, Cho}. The proofs use sophisticated methods 
such as Floer cohomology, and we have no intention to comment on them here. We will 
however need some facts concerning intersecting submanifolds. The dimension of the 
submanifolds is one half times the (real) dimension of the ambient space. 

Consider two circles embedded in a sphere. They are said to {\it intersect transversally} at a 
common point if their tangent vectors at this point span the tangent space of the sphere. 
The sphere has a canonical orientation. If the embedded circles are oriented and ordered 
we can assign an 
index $+1$ or $-1$ to each intersection point depending on whether, at this point, a 
positively directed tangent vector of the first circle followed by a positively 
directed tangent vector of the second circle constitute a positively or negatively 
oriented basis in the tangent space 
of the sphere. The {\it intersection number},  $I$, of the two embedded circles is 
then defined to be the sum of these indices.
One can show that the intersection number is invariant under  
any smooth deformation of the circles. 
See e.g. ref. \cite{Guillemin}.
Now, it is intuitively obvious that the embedded  
circles can always be deformed so that they do not intersect at all.
Therefore $I=0$, which means that the indices assigned to the transversal 
intersection points must vanish in oppositely indexed pairs. 

The intersection number of two Clifford tori in ${\bf CP}^n$ is similarly defined. 
See section 5. Because it is always possible to smoothly deform the tori so 
that they do not intersect, their intersection number must equal zero. 
If, however, we restrict ourselves to Hamiltonian deformations the theorems say 
that it is impossible to completely separate the Clifford tori---at least $2^n$ 
intersection points will stay in the generic case. We will eventually give the 
calculation of the indices in detail, and we will also be concerned with 
how they distribute themselves over the MUB intersection points.

Finally we note that, 
as a glance at Table \ref{tab:states} shows, the theorems we have quoted \cite{Biran, Cho} 
do not tell the full story. 
The motivating question is, can they serve as a spring board from which to address the MUB 
existence problem? The question we will actually address is how things work out in 
${\bf CP}^2$.

\vspace{10mm}

{\bf 4. Birkhoff's polytope}

\vspace{5mm}

\noindent We want to study all possible pairs of Clifford tori in ${\bf CP}^2$. 
In this section we will show that up to natural equivalences the set of all such pairs 
can be parametrized by the four dimensional unistochastic subset of Birkhoff's polytope. 
An alternative standard form for the matrices connecting the tori is given in Appendix A. 

We may assume that the first member of our pair consists of all vectors of the form 
(\ref{Cliff1}) with a flat $\vec{p}$. The second member can be obtained by applying 
a unitary transformation to the first, and the question is where the second torus 
intersects the first torus. Thus, given a unitary matrix $U$, we are looking for all 
choices of phases $(\alpha_1,\alpha_2)$ such that 

\begin{equation} \left( \begin{array}{c} z_0 \\ z_1 \\ z_2 \end{array} \right) 
= U\left( \begin{array}{c} 1 \\ e^{i\alpha_1} 
\\ e^{i\alpha_2} \end{array} \right) \hspace{5mm} \Rightarrow \hspace{5mm} 
|z_0|^2 = |z_1|^2 = |z_2|^2 = 1 
\ . \label{equations}
\end{equation}

\noindent This gives three, or in the general case $N$, real equations, but because 
the matrix is unitary there are in fact only  
$N-1$ independent real equations for $N-1$ unknowns (in dimension $N$). It is not easy 
to guess how many solutions there may be.   

It is clear that diagonal unitaries $D$ and permutation matrices $P$ transform the 
computational Clifford torus into itself.  
For our purposes therefore it is enough to consider equivalence 
classes of unitaries such that 

\begin{equation} U \sim PDUD'P' \ . \end{equation}

\noindent Members of the same equivalence class will give rise to the same 
number of intersection points. Thus we may present all unitaries in  
ordered dephased form, meaning that their first row and first column will be filled 
with non-negative real numbers in non-decreasing order. This equivalence relation arises in 
many contexts \cite{Haagerup}. In practice it is convenient not to impose the permutation 
equivalences explicitly. 

By definition, a {\it bistochastic} (or doubly stochastic) matrix is a square matrix with 
positive entries, such that all rows and all columns sum to one. 
From a unitary matrix we can always construct a bistochastic matrix 
by taking the Hadamard product $U\circ U^* = B$, 

\begin{equation} U \circ U^* = 
\left[ \begin{array}{ccc} |U_{00}|^2 & |U_{01}|^2 & 
|U_{02}|^2 \\ |U_{10}|^2 & |U_{11}|^2 & |U_{12}|^2 \\ 
|U_{20}|^2 & |U_{21}|^2 & |U_{22}|^2\end{array} \right] = 
\left[ \begin{array}{ccc} B_{00} & B_{01} & B_{02} \\ 
B_{10} & B_{11} & B_{12} \\ B_{20} & B_{21} & B_{22} 
\end{array} \right]  = 
B \ . \end{equation}

\noindent Bistochastic matrices arising in this way are called 
{\it unistochastic}, and they form a proper subset of the set of 
all bistochastic matrices. If $U$ is an orthogonal matrix $B$ is said to be 
{\it orthostochastic}. The set of all dephased  unitaries can be 
parametrized by the independent moduli of the matrix elements. 
See ref. \cite{Tadej} for a discussion with full references.   
For $N = 3$ 
there are 4 parameters, and they determine the dephased 
unitary uniquely up to complex conjugation, which again is an operation 
that transforms the computational Clifford torus into itself. 
For higher values of $N$ this parametrization fails, in principle mildly \cite{Karabegov},
in practice badly because the unistochastic subset is difficult to 
characterize. 

The set of all bistochastic matrices ${\cal B}$ is the convex cover of 
the permutation matrices, and is called Birkhoff's polytope \cite{Birkhoff}. 
The nice thing about this, given that we restrict ourselves 
to $N = 3$, is that this set is simple enough to be visualized. Visualizing the unistochastic 
subset ${\cal U}$---which is what we are really interested in---takes a bit more effort, 
but when $N = 3$ we are greatly helped by the fact that in this case its boundary is easy 
to characterize: it consists of the set of all orthostochastic matrices, that is the 
set of bistochastic matrices of the form $B = O\circ O$ where $O$ is an orthogonal 
matrix. Thus we have an analytic description of the boundary. 

Let us give some details (taken from the literature \cite{Tadej}), assuming that 
we have imposed the Euclidean distance 

\begin{equation} D^2(B_1,B_2) = \frac{1}{2}\mbox{Tr}\left[ (B_1-B_2)(B_1-B_2)^{\rm T}\right] \ . 
\end{equation}

\noindent For any bistochastic matrix $B$ there are permutation matrices $P_i$ and a 
probability vector $\vec{\pi}$ such that

\begin{equation} B = \sum_{i=0}^5\pi_iP_i = \left[ \begin{array}{ccc} 
\pi_0 + \pi_1 & \pi_2 + \pi_3 & \pi_4 + \pi_5 \\
\pi_2 + \pi_4 & \pi_0 + \pi_5 & \pi_1 + \pi_3 \\
\pi_3 + \pi_5 & \pi_1 + \pi_4 & \pi_0 + \pi_2 \end{array} \right] \ . \label{permlabel}
\end{equation}

\noindent (Note that this equation provides a labelling of the permutation matrices.) 
The nine facets of the polytope are obtained by setting one matrix element 
equal to zero. The polytope is elegantly described as the convex cover of two totally 
orthogonal regular triangles obtained by setting either $\pi_0=\pi_3=\pi_4 = 0$ 
or $\pi_1=\pi_2 =\pi_5 = 0$. Its centre of mass sits at the van der Waerden matrix 
$B_\star$, a bistochastic matrix with all entries equal. 
The relative volume of the unistochastic subset is \cite{Dunkl} 

\begin{equation} \frac{\mbox{Vol}[{\cal U}]}{\mbox{Vol}[{\cal B}]} = 
\frac{8\pi^2}{105} \approx 0.752 \ . \end{equation} 

\noindent Its insphere has radius $r_{\rm in} = 1/3$, and its outsphere (which coincides 
with the outsphere of Birkhoff's polytope) has radius 1. Moreover ${\cal U}$ is star 
shaped, i.e. fully visible from its centre.  

\begin{figure}
\centerline{ \hbox{
                \epsfig{figure=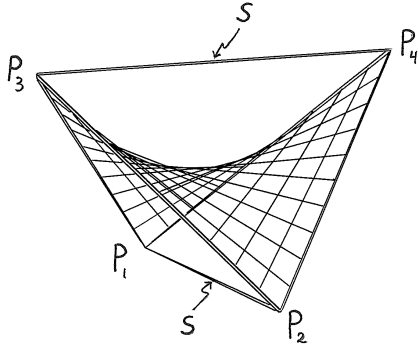,width=60mm}}}
\caption{\small{A facet of Birkhoff's polytope. Four edges are unistochastic 
and bound a ruled surface of unistochastic matrices. Two longer edges 
are outside the unistochastic set. The matrices at the midpoints 
of the longer edges are called Schur matrices and are labelled $S$.}} 
\label{fig:toripap1}
\end{figure}

To see all this in front of us, we begin by looking at a facet 
of Birkhoff's polytope. 
It is a tetrahedron 
with four edges of length $\sqrt{2}$ and two edges of length $\sqrt{3}$. The unistochastic 
subset meets the facet in a two dimensional ruled surface, coming from a maximal torus 
in the group manifold of $SO(3)$. The short edges are unistochastic, the 
long ones not. See Fig. \ref{fig:toripap1}. 

\begin{figure}
\centerline{ \hbox{
                \epsfig{figure=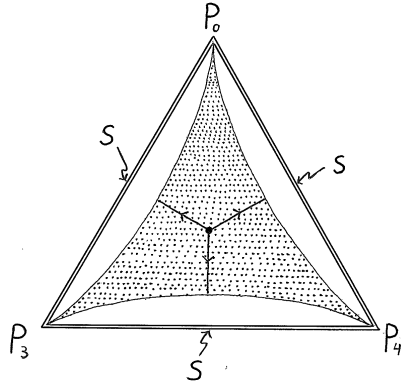,width=55mm}}}
\caption{\small{A cross section through Birkhoff's polytope, centred at the van 
der Waerden matrix and including three non-unistochastic edges. The unistochastic 
subset (shaded) is bounded by a hypocycloid (covering unitaries with exceptional 
properties---see Section 6). Its insphere touches this boundary right below 
the Schur matrices ($S$). The paths emerging from the centre reappear as paths ``b'' in 
Fig. \ref{fig:toripap5}. Birkhoff's polytope is 
the convex cover of two totally orthogonal triangles of this type.}} 
\label{fig:toripap2}
\end{figure}

\begin{figure}
\centerline{ \hbox{
                \epsfig{figure=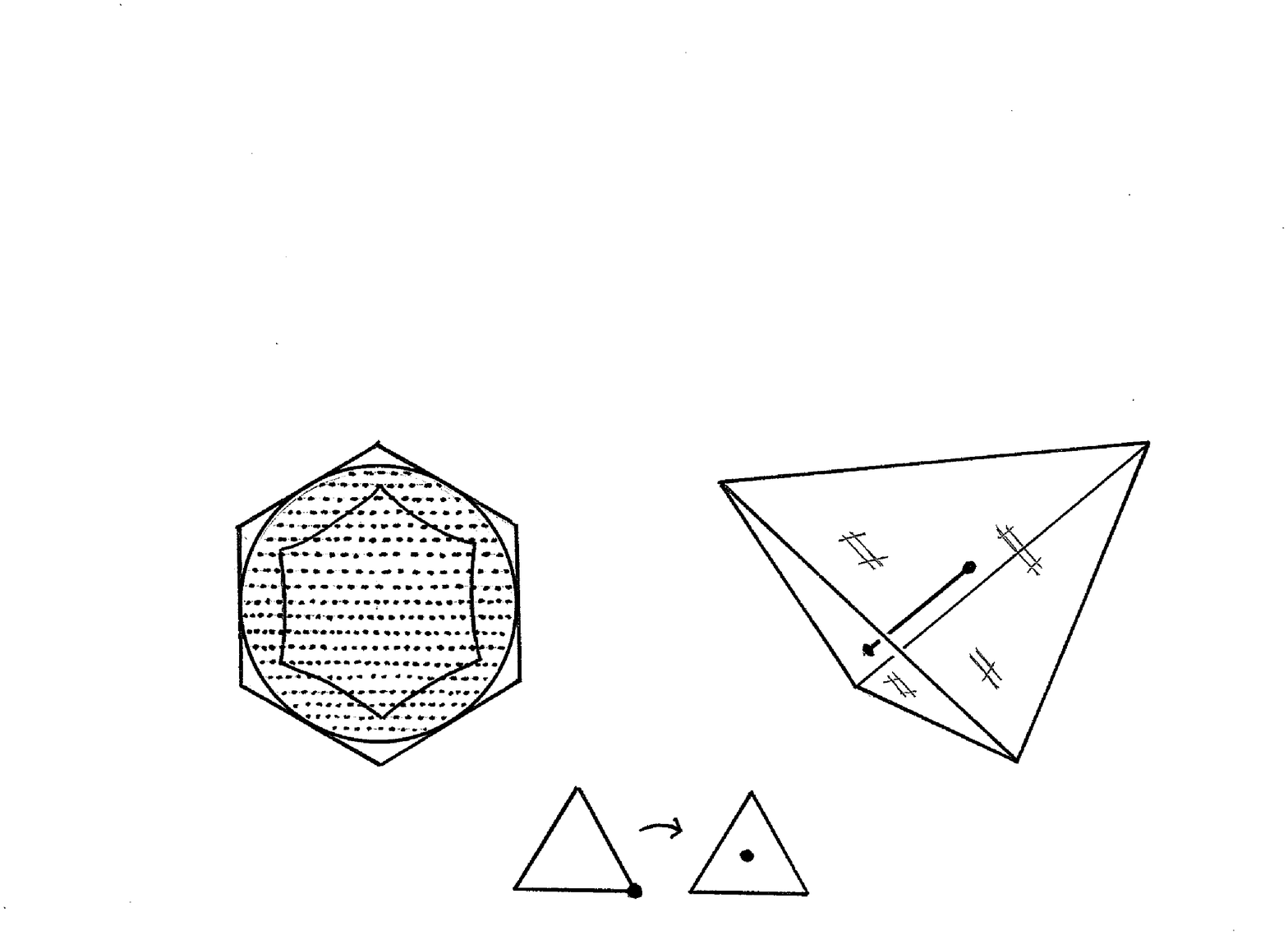,width=95mm}}}
\caption{\small{A hexagonal cross section. Its edges lie in facets, as shown on the right, 
with its corners at midpoints of faces. The unistochastic subset (shaded) is bounded by 
a circle, and its interior is divided into two regions by a boundary to be discussed in section 6.}} 
\label{fig:toripap3}
\end{figure}

\begin{figure}
\centerline{ \hbox{
                \epsfig{figure=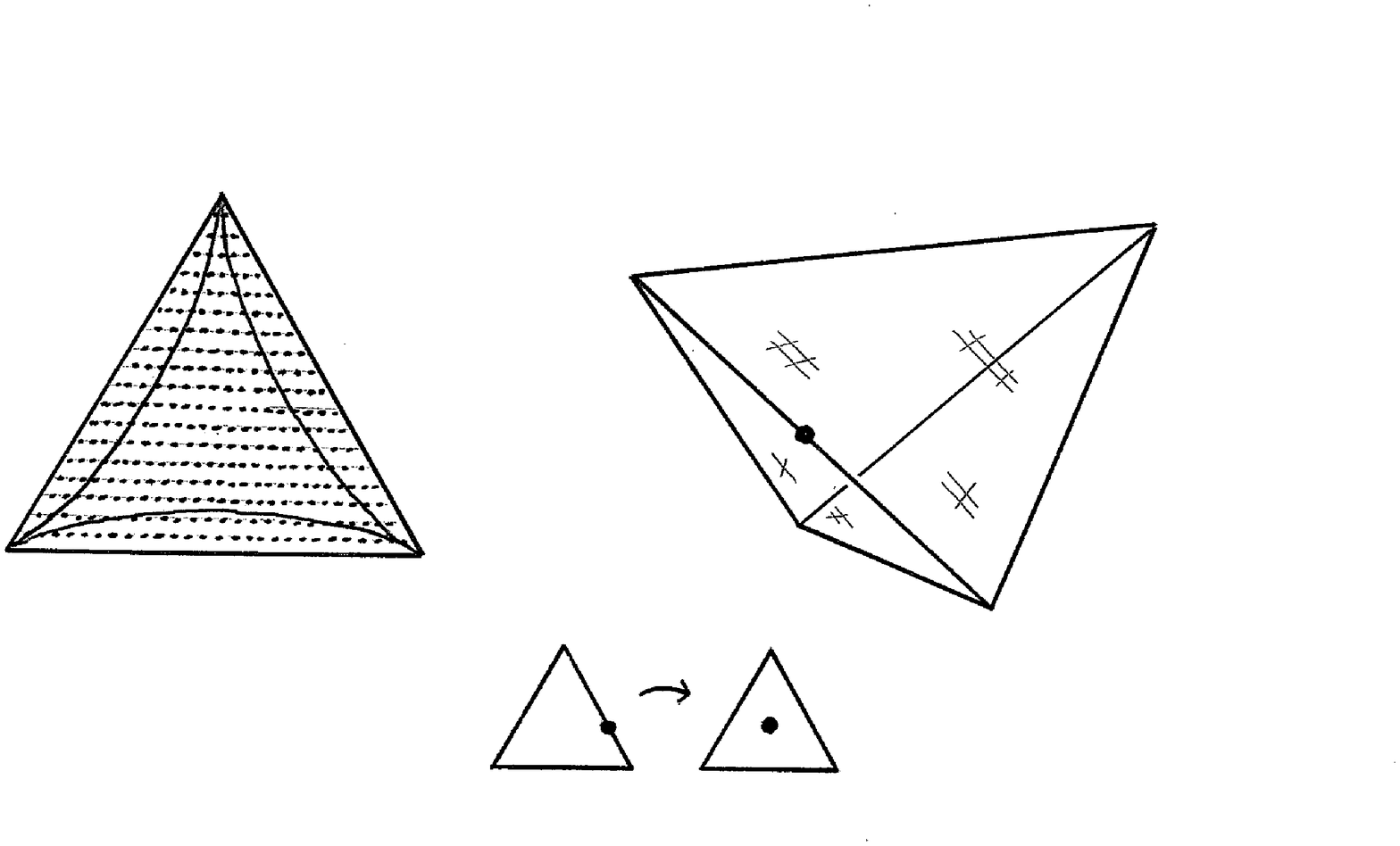,width=95mm}}}
\caption{\small{A triangular cross section, consisting entirely of unistochastic matrices. 
Its corners lie at midpoints of unistochastic edges in the facets, as shown on the right. 
The cross section is divided into two 
regions by a hypocycloid, as discussed in section 6.}} 
\label{fig:toripap4}
\end{figure}

Next we look at a selection of cross sections through the polytope. First we take the 
cross section $\pi_1=\pi_2 =\pi_5 = 0$. This is an equilateral triangle bounded by three 
non-unistochastic edges, and the polytope itself is the convex cover of two such triangles. 
The curve bounding the unistochastic subset is known as a deltoid. The three points on 
this curve closest to the centre lie on the insphere of the unistochastic subset. See Fig. 
\ref{fig:toripap2}. 

Moving somewhat beyond ref. \cite{Tadej}, we also look at cross sections obtained by 
taking all bistochastic matrices 
such that $B\vec{p} = \vec{e}$, where $\vec{p}$ is some fixed probability vector and $\vec{e}$ 
is the flat probability vector. Generically they are hexagons. A case where the hexagon is 
regular is shown in Fig. \ref{fig:toripap3}. Its unistochastic subset is a 
circular disk. A case where the hexagon degenerates to an equilateral triangle is 
shown in Fig. \ref{fig:toripap4}. It lies entirely within the 
unistochastic subset. In both pictures the unistochastic subset is divided into two 
regions, for reasons that will become clear in section 6. 

\begin{figure}
\centerline{ \hbox{
                \epsfig{figure=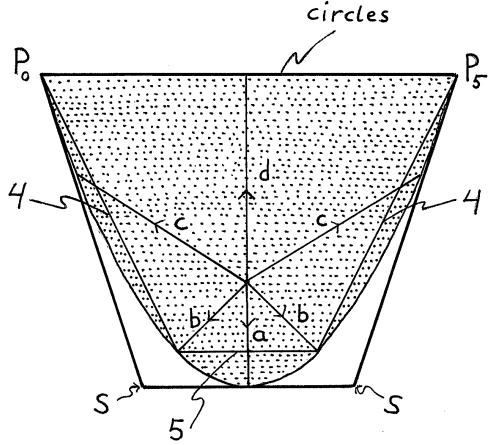,width=70mm}}}
\caption{\small{A cross section including a unistochastic edge and the van der 
Waerden matrix. The unistochastic subset is bounded by a parabola. The region 
where six intersections occur is bounded by an inner boundary consisting of 
straight lines. On these the number of intersections drops to either 4 or 5, as indicated 
by the respective labels. Notice the six paths (of four types) from the van der 
Waerden matrix to the 
boundary. They are displayed in a different way in Fig. \ref{fig:toripap7}. Paths of 
type ``b'' also appear in Fig. \ref{fig:toripap2}.}} 
\label{fig:toripap5}
\end{figure}

A fourth cross section is obtained by including $B_\star$ and a 
unistochastic edge. The boundary of its unistochastic subset is a parabola. The bisectrices 
of the triangles making up the first kind of cross section occur as diagonals in these 
parabolic cross sections. See Fig. \ref{fig:toripap5} (which is placed in section 
6 because we will provide more details there). 

\vspace{10mm}

{\bf 5. Intersections---at the extremes}

\vspace{5mm}

\noindent We consider pairs of Clifford tori related by a unitary matrix according to 
eq. (\ref{equations}). Using the parametrization of the pairs given in section 4 we begin 
with pairs whose parameterizing unitaries sit inside a facet of Birkhoff's polytope. 
In each facet these unitaries make up a 
two dimensional set of orthogonal matrices, and it is easy to verify analytically that 
they relate tori intersecting in exactly 4 points, with the exception of those 
matrices that sit on an edge of the polytope \cite{FR}. The latter relate tori that intersect 
non-transversally in two non-intersecting circles. For an example see panel ``d'' in 
Fig. \ref{fig:toripap7}. If the matrix is a permutation matrix the tori coincide.

Then we venture inside, and choose the unitary to equal the Fourier matrix $F$, 

\begin{equation} F = \frac{1}{\sqrt{3}}
\left( \begin{array}{ccc} 1 & 1 & 1 \\ 1 & \omega & \omega^2 \\ 
1 & \omega^2 & \omega \end{array} \right) \ , \hspace{8mm} \omega = e^{\frac{2\pi i}{3}} 
\ . \end{equation}

\noindent Its columns form the Fourier basis. The corresponding bistochastic matrix 
is the van der Waerden matrix, and sits at the centre of Birkhoff's polytope. 
It is known that the computational and Fourier basis can be supplemented with two 
additional bases to form a complete set of four mutually unbiased bases \cite{Ivanovic}. 
This means that there exist (at least---and in fact exactly) six vectors unbiased 
to both, and hence the two Clifford tori intersect in six points. The intersections 
are transversal, meaning that the 2-dimensional tangent spaces of the two tori span 
the 4-real-dimensional tangent space of the ambient space at the intersections. 
A special feature is that the vectors in the Fourier basis sit in the 
computational Clifford torus. Conversely the computational basis vectors sit in 
the Clifford torus defined by the Fourier basis. 

Now a torus is an orientable manifold, and ${\bf CP}^2$ is oriented by the choice of 
the symplectic form. Following section 3 we can assign an index $\pm 1$ to each 
point where the tori intersect transversally. We permit 
ourselves to use some standard notation from differential geometry, where tangent vectors 
appear as differential operators. We use affine coordinates for the calculation (which 
is given in more
detail in the appendix). Vectors in Hilbert space are written as $(1,z_1,z_2)$. The first 
Clifford torus is described by $(z_1,z_2) = (e^{i\nu_1}, e^{i\nu_2})$, and its tangent 
spaces are spanned by the vector fields 

\begin{equation} \partial_{\nu_1} = iz_1\partial_{z_1} - i \bar{z}_1\partial_{\bar{z}_1} 
\hspace{6mm} \partial_{\nu_2} = iz_2\partial_{z_2} - i \bar{z}_2\partial_{\bar{z}_2}  
\ , \end{equation}

\noindent which are the pushforwards of the coordinate vector fields of $(\nu_1,\nu_2)$, 
computed by the chain rule. The second torus is related to the first by the Fourier 
matrix, and its embedding 
is defined by 

\begin{equation} 
\left( \begin{array}{c} 1 \\ z_1(\alpha_1,\alpha_2) \\ z_2(\alpha_1,\alpha_2)
\end{array} \right) \sim 
\frac{1}{\sqrt{3}}\left( \begin{array}{ccc} 1 & 1 & 1 \\ 
1 & \omega & \omega^2 \\ 1 & \omega^2 & \omega \end{array} \right) \left( 
\begin{array}{c} 1 \\ e^{i\alpha_1} \\ e^{i\alpha_2} \end{array} \right) \ . 
\label{16} \end{equation}

\noindent The $\sim$ symbol means equality up to an overall complex number. In this 
way we obtain a parametric description of the second torus. Finally we 
calculate the pushforwards of the coordinate vector fields under (\ref{16}), 
and call them $\partial_{\alpha_1}, \partial_{\alpha_2}$. We then have bases in 
the tangent spaces of the tori, which we declare to be orientation determining. 
The tori intersect at the six points 

\begin{equation} \left( \begin{array}{c} z_1 \\ z_2 \end{array} \right) 
= \left[ \begin{array}{cccccc} \omega^2 & 1 & \omega & \omega & \omega^2 & 1 \\ 
\omega^2 & \omega & 1 & \omega & 1 & \omega^2 \end{array} \right] \ . \end{equation} 

\begin{figure}
\centerline{ \hbox{
                \epsfig{figure=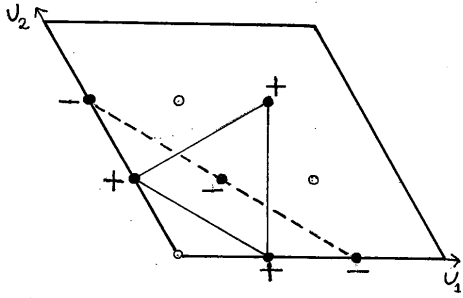,width=65mm}}}
\caption{\small{The computational Clifford torus is drawn with its correct shape as 
given by eq. (\ref{torus}). Opposite sides are identified. The 6 points where this torus  
is intersected by a second torus, related to the first by the Fourier matrix, are 
shown with their 
indices. They form two bases, as indicated (due to the periodicity the dashed line 
is another triangle in disguise). The four bases involved in this construction form 
a complete set of MUB.}} 
\label{fig:toripap6}
\end{figure}

\begin{figure}
\centerline{ \hbox{
                \epsfig{figure=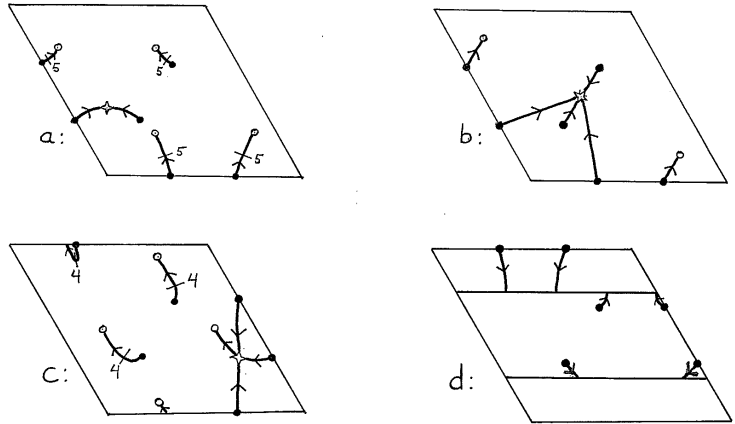,width=100mm}}}
\caption{\small{How the intersections points move and merge as we 
change the matrix from Fourier to the boundary along the four different paths 
shown in Fig. \ref{fig:toripap5}. In case b (also shown in Fig. \ref{fig:toripap2}) one 
of the bases remains a basis all along the path---and it could have 
been kept fixed, see Appendix A. In case d we approach a pair of Clifford tori 
intersecting in two circles. The behaviour is ruled by the indices assigned in Fig. 
\ref{fig:toripap6}, and the coordinates used are the $\alpha_i$ from eq. (\ref{equations}).}} 
\label{fig:toripap7}
\end{figure}

\noindent Our altogether four tangent vector fields span the tangent spaces of ${\bf CP}^2$ 
at these points, and we can calculate the orientation of the bases there. The end 
result of this somewhat lengthy calculation is that 

\begin{equation} \det{ \left[ \begin{array}{cccc} \partial_{\nu_1} & 
\partial_{\nu_2} & \partial_{\alpha_1} 
& \partial_{\alpha_2} \end{array}\right] } = \pm 3 \ , \end{equation}

\noindent with the plus sign occurring three times and the minus sign three times. Thus the 
intersection number $I = 0$, as advertized in section 3. But the calculation also gives 
an index associated to each of the intersections, and therefore also some information 
about how the intersection points can merge, or not, as we deform the tori. Interestingly 
the signs are all plus for a triplet of points corresponding to one basis, and all 
negative for the other. See Fig. \ref{fig:toripap6}. The question whether this pattern 
repeats itself in all odd prime dimensions is addressed in appendix B.

If we move the second torus by means of a continuous family of unitary transformations 
the location of the intersection points will change, and they may even merge and disappear. 
But the sum of the indices of the intersection points will be preserved during this 
process, and this restricts the pattern of possible mergers. 
We know what goes on at the boundary and at the centre of our parameter space. Except 
for a set of measure zero the boundary consists of unitaries that give rise to only 
4 intersection points. In an open ball around the centre 6 intersection points arise. 
Somewhere in between must lie a boundary between those unitaries relating tori that 
intersect 4 times, and those relating tori intersecting 6 times. The aim of the next 
section is to pin this boundary down.

\vspace{10mm}

{\bf 6. How the intersections move and merge}

\vspace{5mm}

\noindent We want a precise account of how the intersections change as we vary the 
pairs of tori. We begin by looking at pairs related by a matrix picked from the 
cross section in Fig. \ref{fig:toripap5}. Four paths from its centre to its boundary 
are drawn and labelled. At the starting point (the Fourier matrix) there are 6 
intersections, supplied with indices according to Fig. \ref{fig:toripap6}. The 
motion of the intersection points is shown in Fig. \ref{fig:toripap7}. As we move 
along the path labelled ``a'' a pair of intersections merge and disappear when we cross 
the part of the inner boundary marked ``5'' in the picture. The remaining 4 intersections 
continue to move until we hit the boundary of the unistochastic set. 
Along path ``b'' the number of distinct 
intersections stays 6 all the way to the corner, where it drops to 3. Along path 
``c'' three intersections merge when we reach the inner boundary marked ``4'' in the picture. 
On the other side of this boundary the number of intersections is 4. Along path ``d'' the 
number of intersections remains 6 all the way to the unistochastic edge, where the 
intersections form two circles. The behaviour is qualitatively the same for all paths from 
the centre, the one thing that matters is which part of the boundary---``5'', ``4'', or 
neither---that is being crossed.

Inspection of Fig. \ref{fig:toripap7}b suggests that it may be possible to continuously 
and unitarily move the Clifford torus of the Fourier basis into coincidence with that 
of the computational basis in such a way that three intersection points (belonging to a 
single basis) remain fixed. We prove this, and sketch the corresponding proof for $N$ 
intersection points in all odd prime dimensions $N$, in Appendix A. For $N = 2$ a look 
at the Bloch sphere suffices as a proof.

The full picture is drawn with limited analytical support. Using 
Mathematica we checked a fairly large number of paths, until we were confident about the 
analytic form of the inner boundary. Then we checked numerically that matrices along this 
inner boundary behave as expected, 
and finally we calculated the intersections analytically in a few examples. 

Which of these behaviours is typical? To answer this question we repeated the exercise for the 
other three cross sections we have studied. In the example shown in Fig 
\ref{fig:toripap3} 
the cross section is cleanly divided into two regions with 6 intersections in the central region, 
4 intersections in the exterior region, and 5 intersections along the boundary. In Fig. 
\ref{fig:toripap4} the behaviour is similar, except if the path escapes at the corners where 
the intersections grow into two circles. There are 5 intersections along the inner boundary, 
which forms a hypocycloid. The cross section shown in Fig. \ref{fig:toripap2} is a 
very special case. In this case there are 6 intersections throughout the interior, and their 
number drops to 3 all along its boundary (except, of course, at the corners). This is not in 
contradiction to the theorems \cite{Biran, Cho} since the intersections cease to be transversal 
there. 

Thus pairs of Clifford tori that intersect in only 3 distinct points occur  at 
the boundary of the unistochastic set in the two special cross sections shown in 
Fig. \ref{fig:toripap2}. 
Incidentally, the path ``b'' from Fig. \ref{fig:toripap5} lies along 
a bisectrix in Fig. \ref{fig:toripap2}. This gives us a useful double check on 
the neighbourhood of the interesting matrix encountered at the boundary there. For this 
very symmetrical choice of path one of the bases formed by the intersection points moves as a unit, 
and remains a basis all along the path. The paths ``a'' and ``d'' lie along a bisectrix in 
Fig. \ref{fig:toripap4} and provide a double check on the behaviour at the unistochastic edge. 

\begin{table}
\caption{{\small Number of randomly generated unitary matrices relating Clifford 
tori with \# intersections (out of $10^5$ examples in dimensions 3 and 4).}}
  \smallskip
  \smallskip
\hskip 1.6cm
{\renewcommand{\arraystretch}{1.1}
\begin{tabular}
{|l| c|c|c|c|c|}\hline \hline
Dimension 3 & \# = 4 & \# = 6 & & & \\
 \hline
 & 80 591 & 19 409 & & & \\
\hline \hline 
Dimension 4 & \# = 8 & \# = 10 & \# = 12 & \# = 14 & \# = 16 \\ 
\hline 
 & 31 882 & 44 425 & 22 023 & 1 605 & 65 \\
\hline \hline
\end{tabular}
}
\label{tab:random}
\end{table} 

One of our aims was to find an analytic description of the boundary separating 
regions of matrices giving rise to Clifford tori with 6 and 4 intersections, 
respectively. In this we failed, but the cross sections we have shown give much 
information about its shape. They suggest, as expected, that the boundary between the 
two regions generically consists of matrices with 5 cross sections, although exceptions 
were identified. They support the conjecture that the region giving rise 
to 6 intersections is star shaped. In looking at the pictures we should remember that 
Birkhoff's polytope lives in four dimensions, and most of its volume lies far from 
its centre. When we picked unitary matrices at random using the Haar measure we found 
that the region giving rise to 6 intersections occurs about one fifth of the time (as 
has been independently confirmed \cite{Zbig}). The average Euclidean distance from the centre 
of Birkhoff's polytope was $0.49$ for matrices leading to 4 intersections, and $0.37$ for 
those leading to 6 intersections. Table \ref{tab:random} gives 
precise results. For each matrix generated we solved eq. (\ref{equations}) numerically to 
60 digits precision. The exercise was then repeated for $N = 4$ dimensions in order to 
see if the lowest possible number of intersections is always the most common case. It is not. 

\vspace{10mm}

{\bf 7. Discussion}

\vspace{5mm}

\noindent In conclusion, Clifford tori in ${\bf CP}^1$ always intersect in two points. 
The first non-trivial case occurs in ${\bf CP}^2$. In this case generic pairs of Clifford 
tori intersect in either 4 or 6 points, the former case having the largest Haar measure. 
(In four dimensions the largest measure pertains to Clifford tori intersecting more 
than the $2^3$ times guaranteed by the theorems \cite{Biran, Cho}.) 
We observed that pairs of Clifford tori can be parametrized, up to natural equivalences, 
by the unistochastic subset ${\cal U}$ of Birkhoff's polytope, which in this case is a four 
dimensional set whose structure is well understood. This set is divided into 
two regions. The region corresponding to 6 intersections contains the centre of ${\cal U}$, 
while the closure of the region corresponding to 4 intersections contains the boundary of ${\cal U}$. 
On the boundary between these regions one typically but not 
always finds 5 distinct intersections. Clifford tori intersecting in only 3 points 
occur along two distinct curves in the set of inequivalent pairs, namely on the boundary of the 
unistochastic set in the two special cross sections depicted in Fig. \ref{fig:toripap2}.

We described the situation by giving a precise account of four different two dimensional 
cross sections of ${\cal U}$. Based on them we conjecture that the 
region corresponding to 6 intersections 
is star shaped, and that pairs intersecting in just 3 distinct points occur only along 
the two curves we identified.

The intersection number for a pair of Clifford tori vanishes, but the individual signs 
distribute themselves in an interesting way when there are six intersection points 
corresponding to a MUB: the signs are all plus for one basis, and all negative for the 
other. This pattern recurs in (at least) all odd prime dimensions $\leq 17$, so this is 
not an $N = 3$ coincidence. It is 
connected to the fact that the MUB 
vectors fall in two groups, depending on whether their components can be constructed 
from the root of unity $e^{2\pi i/N}$ raised to quadratic residues or quadratic non-residues.

Now the question is whether we have learned anything of use for the MUB existence problem? 
In Figs. \ref{fig:toripap2} and \ref{fig:toripap5} the arrows clearly go the ``wrong'' way. 
The calculations were performed along curves starting at the Fourier case and moving outwards 
towards the boundary of the set. What we would really like to do is to set up some kind of 
natural flow in the set of pairs of Clifford tori, such that the intersection pattern 
changes towards the pattern that signals the presence of a complete set of MUBs. In dimensions 
where the MUB existence problem has a negative answer, obstructions to this flow should be 
identified. Whether anything like this can be achieved we do not know.

It will occur to the reader that, since we restricted ourselves to the lowest non-trivial 
dimension, our pictures do not even begin to capture the complexities hidden behind Table 
\ref{tab:states}. This is true. Of course there are contexts in which the qutrit, or 
${\bf CP}^2$, is of special interest. 
Moreover, the geometry of the unistochastic subset of 3 by 3 bistochastic matrices is 
of independent interest \cite{Turok}. Nevertheless, our purpose was to inject new ideas 
that may lead to progress in all dimensions, and we hope they will. 

\

\

\noindent \underline{Acknowledgements}

\

\noindent We thank Karol \.Zyczkowski for drawing our attention to Cho's work, 
Zbigniew Pucha{\l}a for sending nice pictures, 
and David Andersson for technical help.

\

\

{\bf Appendix A: Some calculations in higher dimensions}

\vspace{5mm}

\noindent We have characterized pairs of Clifford tori by the unitary matrices connecting 
them, and we divided the latter into equivalence classes, choosing a dephased unitary as 
representing each class. This is very convenient in 3 dimensions, but in higher dimensions 
an alternative standard form may be preferable. Given the theorems on which we rely 
\cite{Biran, Cho} we know that every Clifford torus intersects the computational Clifford torus, 
and by translations along the tori (that is, by enphasing the unitary that connects them) 
we can ensure that one of the intersection points is represented by the flat vector (all of whose 
entries are equal to $1/\sqrt{N}$). Denote this vector by $|f_0\rangle$, and extend 
it to an orthonormal basis $\{ |f_i\rangle \}_{i=0}^{N-1}$ in some way. Then each equivalence 
class has a representative unitary matrix of the form

\begin{equation} U = |f_0\rangle \langle f_0| + \sum_{i=1}^{N-1}e^{i\sigma_i}|f_i\rangle 
\langle f_i| \ . \end{equation}

\noindent By construction, the flat vector is left invariant by this matrix. There are 
$(N-1)(N-2)$ real parameters involved in the choice of the basis 
and $N-1$ parameters in the eigenvalues, so the dimension of the set of inequivalent 
pairs of Clifford tori is $(N-1)^2$, which is equal to the dimension of Birkhoff's 
polytope.  

We can now give an example of a one-parameter family of unitaries $U(\sigma )$, $\sigma 
\in [0,\pi]$, interpolating 
between the identity and a matrix in the same equivalence class as the Fourier matrix $F$, 
and which relates Clifford tori that intersect at the vectors in the Fourier basis for all 
values of the parameter. The matrix elements are

\begin{equation} U_{uv}(\sigma) = \frac{1}{N}\sum_{s=0}^{N-1}e^{i\sigma s^2}
\omega^{(u-v)s} \ , \hspace{6mm} 0 \leq u, v \leq N-1 \ , 
\hspace{6mm} \omega = e^{\frac{2\pi i}{N}} \ .  \end{equation}

\noindent This substantiates the claim made in section 5, namely that in all prime dimensions 
it is possible to continuously and unitarily move the standard torus into coincidence with the 
Fourier torus in such a way that $N$ intersection points, forming one of the bases in a MUB, 
are left fixed. 

The proof uses a standard Gauss sum valid for odd prime $N$, eq. (\ref{gaussum}) below. 
However, the idea becomes much 
more transparent if we write it out for $N = 3$, in which case the one-parameter family is 
given by 

\begin{equation} U(\sigma) = \frac{1}{3}\left( \begin{array}{c} 1 \\ 1 \\ 1 \end{array}\right) 
( 1 \ 1 \ 1) + \frac{e^{i\sigma}}{3}\left( \begin{array}{l} 1 \\ \omega \\ \omega^2 
\end{array}\right) 
( 1 \ \omega^2 \ \omega ) + \frac{e^{i\sigma}}{3}\left( \begin{array}{l} 1 \\ \omega^2 
\\ \omega \end{array}\right) 
( 1 \ \omega \ \omega^2) \ . \end{equation}

\noindent For $\sigma = 0$ this gives the identity, for $\sigma = 2\pi /3$ it covers 
the van der Waerden matrix, and for $\sigma = \pi$ it reaches the boundary of the 
unistochastic set. The six intersection points of the tori related by this matrix are 
given by vectors $(1, e^{i\alpha_1},e^{i\alpha_2})$ equal to one of the columns of the matrix 

\begin{equation} \left[ \begin{array}{cccccc} 1 & 1 & 1 & 1 & 1 & 1 \\ 
1 & \omega & \omega^2 & - e^{ia} & - e^{ia} & 1 \\ 
1 & \omega^2 & \omega & - e^{ia} & 1 & -e^{-ia} \end{array} \right] \ , \end{equation}

\noindent where 

\begin{equation} e^{ia} = \frac{1+2e^{i\sigma}}{2+e^{i\sigma}} \ . \end{equation}

\noindent This provides an analytic check on Fig. \ref{fig:toripap7}b, although in that picture 
the ``fixed basis'' moves because we used a different standard form of the unitaries. (By the 
way, our figures rely not only on numerical calculations but also on quite a few analytic checks 
of this kind.) 

\vspace{10mm} 

{\bf Appendix B: Calculating the intersection indices}

\vspace{5mm}

\noindent 

\noindent  We will now calculate the intersection indices, in odd prime dimensions, for a pair 
of Clifford tori related by the Fourier matrix $F$. 
In general odd prime dimensions $p$ a complete set of MUB can be taken to consist 
of the computational basis, the Fourier basis, and $p-1$ bases given by columns of circulant 
matrices \cite{Ivanovic}. Let the non-zero integer $z < p$ label the circulant bases 
and the integer $a <p$ 
label the vectors. With $\omega = e^{2\pi i/p}$ and using arithmetic modulo $p$ whenever 
applicable, we have the MUB vectors 

\begin{equation} |z,a\rangle \sim \sum_{r=0}^{p-1}\omega^{\frac{(r-a)^2}{2z}} 
|e_r\rangle \sim \sum_{r=0}^{p-1}\omega^{\frac{r^2-2ra}{2z}}  
|e_r\rangle \ , \label{MUvectors} \end{equation}

\noindent where we removed a phase factor in the last step in order to prepare for using 
affine coordinates. 
When we apply the discrete Fourier transformation the vectors 
are rearranged according to 

\begin{equation} |z,a\rangle \rightarrow |z',a'\rangle = \left| - \frac{1}{z}, 
\frac{a}{z}\right\rangle \ . \end{equation}

\noindent
We use the angles $\nu_r$ to coordinatize 
the fixed torus and the angles $\alpha_r$ to coordinatize the torus that we shift with the 
Fourier matrix. At the intersection points defined by these vectors, see eq. (\ref{MUvectors}), 
we have then that 

\begin{equation} e^{i\alpha_r} = \omega^{\frac{r^2-2ra}{2z}} \hspace{5mm} \Rightarrow \hspace{5mm} 
z_r = e^{i\nu_r} = \omega^{-\frac{zr^2}{2} + ra} \ . \label{punkt0} \end{equation}

\noindent This gives the affine coordinates $z_r$ evaluated at these intersection points. 
A general point on the shifted torus ends up at 

\begin{equation} z_r = z_r(\alpha) = \frac{\sum_{s=0}^{p-1}\omega^{rs}e^{i\alpha_s}}
{\sum_{s=0}^{p-1}e^{i\alpha_s}} \hspace{5mm} \Rightarrow \hspace{5mm} 
\frac{\partial z_s}{\partial \alpha_r} = ie^{i\alpha_r}\frac{\omega^{sr}-z_s}
{\sum_se^{i\alpha_s}} \ . \label{punkt2} \end{equation}

\noindent Finally we can write down the tangent vectors of the two tori at the intersection 
points, namely 

\begin{equation} \partial_{\nu_r} = iz_r\partial_{z_r} - i\bar{z}_r\partial_{\bar{z}_r} 
\end{equation}

\begin{equation} \partial_{\alpha_r} = \sum_{s=0}^{p-1}\left( 
\frac{\partial z_s}{\partial \alpha_r}\partial_{z_r} + \frac{\partial \bar{z}_s}
{\partial \alpha_r}\partial_{\bar{z}_r} \right) \ , \end{equation}

\noindent where the premise in (\ref{punkt0}) is assumed to hold. 

Unfortunately we were unable to calculate the resulting determinant in general. For 
$p = 3, 7, 11$ we found

\begin{equation} \det [ \partial_{\nu_1}, \dots \partial_{\nu_{p-1}},\partial_{\alpha_1}, 
\dots , \partial_{\alpha_{p-1}}] = \pm p \ , \end{equation}

\noindent where the plus sign applies to all vectors coming from a basis $|z,a\rangle$ 
with $z$ a quadratic residue. For $p = 5,13,17$ it is again true that the determinant is positive 
for all vectors coming from a basis whose $z$ is a quadratic residue, but for these three 
cases we found respectively that 

\begin{equation} \det 
= \frac{5}{2} (\pm 3 - \sqrt{5}) \ , \hspace{3mm} \det = \frac{13}{2} 
(\pm 11 - 3\sqrt{13}) \ , \hspace{3mm} \det = 17(\pm 33 - 8\sqrt{17}) \ . \end{equation}

\noindent The difference between primes equal to 3 modulo 4 and primes equal to 1 modulo 4 
is due to the Gauss sum 

\begin{equation} \sum_{x=0}^{p-1}  \omega^{x^2} = 1 + 2\sum_{x \in {\bf Q}}\omega^x = 
\left\{ \begin{array}{lll} \sqrt{p} & 
\mbox{if} & p = 4k+1 \\ \\ i\sqrt{p} & \mbox{if} & p = 4k+3 \ , \end{array} \right. 
\label{gaussum} \end{equation}

\noindent where ${\bf Q}$ denotes the set of quadratic residues (i.e., the set of non-negative 
integers that can be written as the square of another integer in arithmetic modulo $p$). Sums 
over roots of unity with quadratic (non-)residues are easily calculated from this formula. 

The conclusion is that we have some evidence for the conjecture that the intersection 
indices are always equal for intersection points taken from the same mutually unbiased 
basis if the dimension is an odd prime---but we did not prove it.


{\small

\begin{thebibliography}{99}

\bibitem{DEBZ} T. Durt, B.-G. Englert, I. Bengtsson, and K. \.Zyczkowski, {\it On 
mutually unbiased bases}, Int. J. Quantum Inf. {\bf 08}, 535, 2010. 

\bibitem{Lam} C. W. H. Lam, {\it The search for a finite projective plane of 
order 10}, Amer. Math. Mon. {\bf 98}, 305, 1991. 

\bibitem{Paterek} T. Paterek, M. Paw{\l}owski, M. Grassl and {\v{C}}. Brukner,
{\it On the connection between mutually unbiased bases and orthogonal Latin squares},
Phys. Scr. {\bf 2010}, 014031, 2010.

\bibitem{Weigert} S. Weigert and T. Durt,
{\it Affine constellations without mutually unbiased counterparts},
J. Phys. A: Math. Theor.
{\bf 43}, 402002, 2010.

\bibitem{Weiner} M. Weiner, {\it A gap for the maximum number of mutually unbiased 
bases}, Proc. Amer. Math. Soc. {\bf 141}, 1963, 2013. 

\bibitem{Arnold} V. I. Arnold: {\it Mathematical methods in classical mechanics}, 
Springer, New York 1989.

\bibitem{Biran} P. Biran, M. Entov, and L. Polterovich, {\it Calabi quasimorphisms for 
symplectic ball}, Commun. Contemp. Math. {\bf 06}, 793, 2004.

\bibitem{Cho} C. H. Cho, {\it  Holomorphic disks, spin structures and Floer cohomology of 
the Clifford torus}, Int. Math. Res. Not. {\bf 2004} 1803.

\bibitem{Schwinger} J. Schwinger: {\it Quantum Mechanics. Symbolism of 
Atomic Measurements}, ed. by B.-G. Englert, Springer, Berlin 2001.

\bibitem{Ivanovic} I. D. Ivanovi\'c, {\it Geometrical description of state determination}, 
{\it J. Phys.} {\bf A14}, 3241, 1981.

\bibitem{Wootters} W. K. Wootters, {\it A Wigner-function formulation of finite-state 
quantum mechanics}, Ann. Phys. {\bf 176}, 1, 1987.

\bibitem{Mo} N. J. Cerf, M. Bourennane, A. Karlsson, and N. Gisin, {\it Security of quantum 
key distribution using d-level systems}, Phys. Rev. Lett. {\bf 88}, 127902, 2002.

\bibitem{Mafu} M. Mafu et al., {\it Higher-dimensional orbital angular momentum 
based quantum key distribution with mutually unbiased bases}, 
Phys. Rev. {\bf A88}, 032305, 2013.

\bibitem{Kostrikin} A. I. Kostrikin and P. I. Tiep: {\it Orthogonal decompositions and 
integral lattices}, de Gruyter, Berlin 1994.

\bibitem{Fields} W. K. Wootters and B. D. Fields, {\it Optimal state-determination by 
mutually unbiased bases}, Ann. Phys. NY {\bf 191}, 363, 1989.

\bibitem{Godsil} C. Godsil and A. Roy, {\it Equiangular lines, mutually unbiased bases, 
and spin models}, Eur. J. Comb. {\bf 30}, 246, 2009.

\bibitem{Kantor} W. M. Kantor, {\it MUBs inequivalence and affine planes}, J. Math. Phys. 
{\bf 53}, 032204, 2012.

\bibitem{Vatan} S. Bandyopadhyay, P. O. Boykin, V. Roychowdhury and F. Vatan, {\it A new 
proof for the existence of mutually unbiased bases}, Algorithmica {\bf 34}, 512, 2002.

\bibitem{ACW} M. Aschbacher, A. M. Childs, and P. Wocjan, {\it The limitations of nice 
mutually unbiased bases}, J. Algebra. Comb. {\bf 25}, 111, 2007.

\bibitem{Zhu} H. Zhu, {\it Mutually unbiased bases as minimal Clifford covariant 
2-designs}, Phys. Rev. {\bf A91}, 060301, 2015.

\bibitem{BS} S. Brierley and S. Weigert, {\it Maximal sets of mutually unbiased 
quantum states in dimension six}, Phys. Rev. {\bf A78}, 04312, 2008. 

\bibitem{Jam} P. Jaming, M. Matolcsi, P. M\'ora, F. Sz{\"o}ll{\H{o}}si, and M. 
Weiner, {\it A generalized Pauli problem and an infinite family of MUB-triplets 
in dimension 6}, J. Phys. {\bf A42}, 245305, 2009.

\bibitem{Raynal} P. Raynal, X. L\"u, and B.-G. Englert, {\it Mutually unbiased
bases in dimension six: The four most distant bases}, Phys. Rev. {\bf A83}, 
062303, 2011.

\bibitem{Gosta} G. Bj\"orck and B. Saffari, {\it New classes of finite unimodular 
sequences with unimodular Fourier transforms}, C. R. Acad. Sci., Paris, S\'er. I 
{\bf 320}, 319, 1995.

\bibitem{Haagerup} U. Haagerup, {\it Orthogonal maximal abelian *-subalgebras of the
$n\times n$ matrices and cyclic $n$-roots}, in \emph{Operator Algebras and
Quantum Field Theory, Rome (1996)}, Internat. Press, Cambridge, MA 1997.

\bibitem{Grassl} M. Grassl, {\it On SIC-POVMs and MUBs in dimension 6}, arXiv 
eprint quant-ph/0406175.

\bibitem{Bengt} B. R. Karlsson, {\it Three-parameter complex Hadamard matrices of 
order 6}, Lin. Alg. Appl. {\bf 434}, 247, 2011.

\bibitem{Szollosi} F. Sz{\"o}ll{\H{o}}si, {\it Complex Hadamard matrices of order
6: a four-parameter family}, J. London Math. Soc. {\bf 85}, 616, 2012.

\bibitem{Rudolph} K. Korzekwa, D. Jennings, and T. Rudolph, {\it Operational 
constraints on state-dependent formulations of quantum error-disturbance 
trade-off relations}, Phys. Rev {\bf A89}, 052108, 2014.

\bibitem{Karol} Z. Pucha{\l}a, {\L}. Rudnicki, K. Chabuda, M. Paraniak, and K. 
\.Zyczkowski, {\it Certainty relations, mutual entanglement and non-displacable 
manifolds}, Phys. Rev. {\bf A92}, 012304, 2015.

\bibitem{Idel} M. Idel and M. M. Wolf, {\it Sinkhorn normal form for unitary matrices}, 
Lin. Alg. Appl. {\bf 471}, 76, 2015.

\bibitem{FR} H. F\"uhr and Z. Rzeszotnik, {\it On biunimodular vectors for unitary 
matrices}, Lin. Alg. Appl. {\bf 484}, 86, 2015.

\bibitem{Kibble} T. W. B. Kibble, {\it Geometrization of quantum mechanics}, Commun. Math. 
Phys. {\bf 65}, 189, 1979.

\bibitem{Brody} D. C. Brody and L. P. Hughston, {\it Geometric quantum mechanics}, 
J. Geom. Phys. {\bf 38}, 19, 2001.

\bibitem{Ingemar} I. Bengtsson and K. \.{Z}yczkowski: {\it Geometry of Quantum States}, 
Cambridge University Press, Cambridge, 2006

\bibitem{Guillemin} V. Guillemin and A. Pollack: {\it Differential Topology}, Prentice-Hall, 
New Jersey 1974.

\bibitem{Tadej} I. Bengtsson, \AA . Ericsson, M. Ku\'s, W. Tadej, and K. \.Zyczkowski, 
{\it Birkhoff's polytope and unistochastic matrices, $N = 3$ and $N = 4$}, 
Commun. Math. Phys. {\bf 259}, 307, 2005.

\bibitem{Karabegov} A. V. Karabegov, {\it A mapping from the unitary to the 
doubly stochastic matrices and symbols on a finite set}, AIP Conf. Proc. 
{\bf 1079}, 39, 2008.

\bibitem{Birkhoff} G. Birkhoff, {\it Tres observaciones sobre el algebra lineal}, 
Univ. Nac. Tucum\'an Rev. {\bf A5}, 147, 1946. 

\bibitem{Dunkl} C. Dunkl and K. \.Zyczkowski, {\it Volume of the set of unistochastic 
matrices of order 3 and the mean Jarlskog invariant}, J. Math. Phys. {\bf 50}, 
123521, 2009.

\bibitem{Zbig} Z. Pucha{\l}a, private communication.

\bibitem{Turok} G. W. Gibbons, S. Gielen, C. N. Pope, and N. Turok, {\it Measures on 
mixing angles}, Phys. Rev. {\bf D79}, 013009, 2009.




\end{thebibliography}

}
\end{document}